\documentclass[prl,twocolumn,citeautoscript,superscriptaddress,floatfix]{revtex4}

\usepackage{hyperref}
\usepackage{grffile}
\usepackage{graphicx}
\usepackage{amsmath}
\usepackage{cases}
\usepackage{amssymb}
\usepackage{color}
\usepackage{mathrsfs}
\usepackage{braket}
\usepackage{xr}
\begin{document}

\title{Symmetry-protected transport through a lattice with a local particle loss}

\author{A.-M. Visuri}
\email{avisuri@uni-bonn.de}
\affiliation{Physikalisches Institut, University of Bonn, Nussallee 12, 53115 Bonn, Germany}
\author{T. Giamarchi}
\affiliation{Department of Quantum Matter Physics, University of Geneva, 24 quai Ernest-Ansermet, 1211 Geneva, Switzerland}
\author{C. Kollath}
\affiliation{Physikalisches Institut, University of Bonn, Nussallee 12, 53115 Bonn, Germany}

\begin{abstract}
We study particle transport through a chain of coupled sites connected to free-fermion reservoirs at both ends, subjected to a local particle loss.
The transport is characterized by calculating the conductance and particle density in the steady state using the Keldysh formalism for open quantum systems. Besides  a reduction of conductance, we find that transport can remain (almost) unaffected by the loss for certain values of the chemical potential in the lattice. We show that this ``protected'' transport results from the spatial symmetry of single-particle eigenstates. At a finite voltage, the density profile develops a drop at the lossy site, connected to the onset of non-ballistic transport.
\end{abstract}

\maketitle

Over the decades, dissipative processes have been considered a nuisance in quantum systems since they destroy the essential property -- quantum coherence. This usually has detrimental consequences for applications such as  quantum computing. Recently, this  point of view was revisited and reversed so that dissipative processes are now \emph{employed} as a tool to bring quantum systems into desired states with novel features \cite{MuellerZoller2012}. For example, a dissipative coupling was used to prepare phase- and number-squeezed states with ultracold atoms \cite{CaballarWatanabe2014},
a Tonks-Girardeau gas of molecules \cite{SyassenDuerr2008} and even entanglement among trapped ions \cite{BarreiroBlatt2011}, while environment-assisted quantum transport~\cite{VicianiCaruso2015,MaierRoos2019,DolgirevDemler2020} was demonstrated using dephasing noise.

Dissipation is usually understood as irreversible energy loss due to coupling to an environment. Among dissipative mechanisms, the loss of particles plays an important role. Experiments with cold atoms offer a platform to engineer and study particle losses in a controlled way. One realization was to apply an electron beam to weakly-interacting bosonic gases~\cite{BarontiniOtt2013,LabouvieOtt2016}, 
paving the way to the study of new phenomena. Theoretically, the influence of a localized loss or dephasing has been studied extensively for weakly-interacting bosonic atoms \cite{BrazhnyiOtt2009,TonielliMarino2020} and the Bose-Hubbard model~\cite{BarmettlerKollath2011,WitthautWimberger2011,KieferSirker2017}.
For fermionic systems, less is known, and only recently, a local particle loss was realized in a cold-atom experiment using near-resonant optical tweezers~\cite{CormanEsslinger2019,LebratEsslinger2019}.
Theoretical analyses~\cite{WolffKollath2020,FromlDiehl2019,FromlDiehl2020,MullerDiehl2021} have shown evidence for a quantum Zeno effect~\cite{MisraSudarshan1977,BreuerPetruccione2002}
where the interplay of the interaction and loss pushes the system to a new steady state with peculiar properties, different from the equilibrium ones.

Given the striking effects of particle losses when the system is otherwise in equilibrium, it is important to understand their consequences in a system which is already in a non-thermal steady state. 
Such steady states occur quite generally when a system is coupled to two reservoirs at different 
chemical potentials, leading to the transport of matter~\cite{Landauer1957,NazarovBlanter2009,AkkermansMontambaux2007}. Steady-state transport is one of most common probes of the properties of quantum systems and has been extensively applied in the condensed-matter context. In solid-state junctions, a loss or gain of electrons can be implemented through additional leads -- this technique was applied to controlling supercurrents in Josephson junctions~\cite{MorpurgoVanWees1998,BaselmansKlapwijk1999,MorpurgoKlapwijk2000,CrosserBirge2006}. More recently, particle transport between reservoirs has been studied in cold-atom experiments~\cite{KrinnerBrantut2017}, where the effects of local particle losses on transport were also explored~\cite{CormanEsslinger2019}.
Understanding the consequences of particle losses in a nonequilibrium steady state, as opposed to equilibrium, is thus prompted by recent experiments but is also interesting from a theoretical point of view. This novel situation poses new conceptual problems since it combines two different ways to push the system out of equilibrium. So far, it has been treated by approximate methods, such as incorporating an imaginary potential in the Landauer-B\"uttiker 
formula of transport~\cite{CormanEsslinger2019}, or describing the reservoirs by Lindblad boundary conditions~\cite{DamanetDaley2019,DamanetDaley2019Nov,JinGiamarchi2020}. A full analysis is clearly difficult and yet little explored.

In this paper, we address the effects of losses on the transport through a quantum dot or an extended lattice coupled to two reservoirs. We use a full Keldysh description, allowing for an exact solution. We obtain 
both the conductance of the lossy system and the density profile in presence of losses and a finite voltage. Surprisingly, the conductance can be robust to losses, which we relate to the inversion symmetry in an isolated lattice. We show that for intermediate to large losses, a voltage drop occurs across the dissipative defect in contrast to the ballistic behavior of a lossless system. 

\begin{figure}
\includegraphics[width=\linewidth]{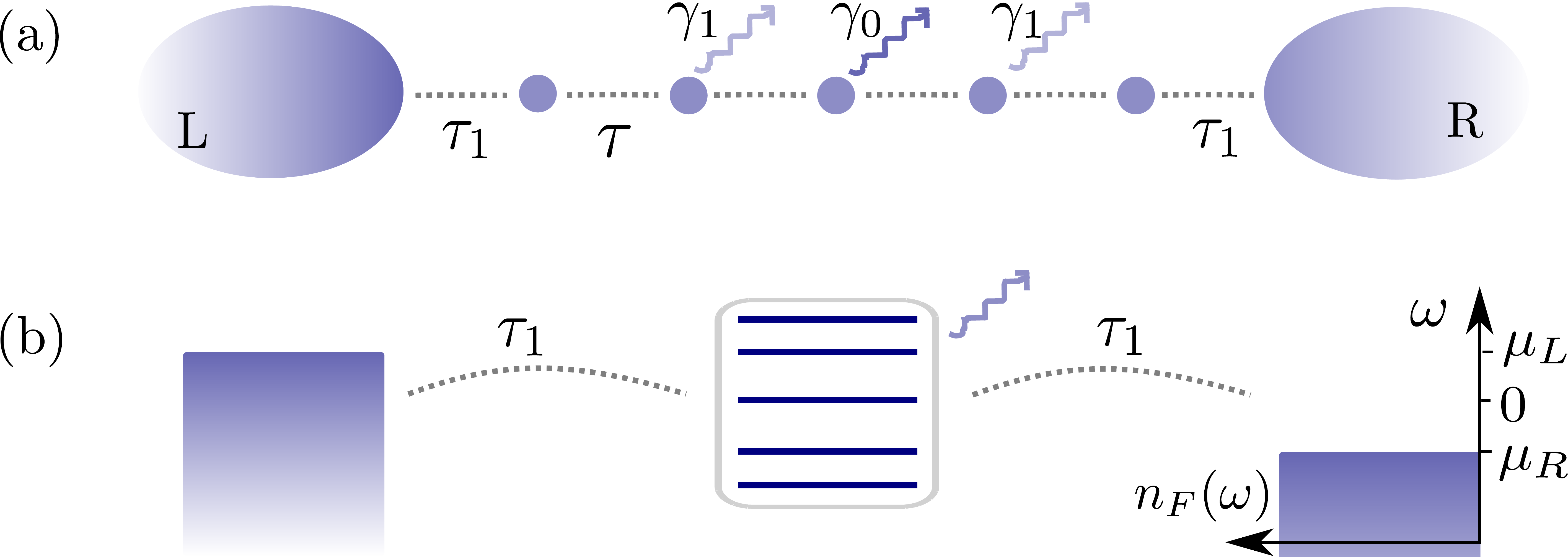}
\caption{(a) A chain of $M$ sites is coupled to reservoirs and subjected to a local particle loss with amplitude $\gamma_0$ on the central site and $\gamma_1$ on the neighboring sites. (b) The Fermi distribution $n_F(\omega)$ at zero temperature: states up to the chemical potentials $\mu_{L, R}$ in the reservoirs are filled. The horizontal lines within the box in the middle show the eigenenergies of an isolated lattice. Resonances in transport occur when an eigenenergy coincides with $\mu_{L, R}$ in the limit $\mu_L-\mu_R \to 0$.}
\label{fig:dissipative_chain}
\end{figure}

We consider a one-dimensional lattice coupled at both ends to a free-fermion reservoir and subjected to a local particle loss at the center (see Fig.~\ref{fig:dissipative_chain}). 
The system without the loss is described by the Hamiltonian
$H = \sum_{i = L, R} H_i + H_{\text{chain}} + H_{\text{t}}$.
The subscripts $L,R$ denote the left and right reservoirs, respectively, with the Hamiltonian  
$H_i = \sum_{\mathbf{k}} \left(\epsilon_{\mathbf{k}} - \mu_i \right) \psi_{i \mathbf{k}}^{\dagger} \psi_{i \mathbf{k}}^{\phantom{\dagger}}$. 
Here, $\mathbf{k}$ denotes a quantum number (typically momentum), 
$\epsilon_{\mathbf{k}}$ the energy, 
and $\psi_{i \mathbf{k}}^{\dagger}$ ($\psi_{i \mathbf{k}}^{\phantom{\dagger}}$) the fermionic creation (annihilation) operator of reservoir $i$. The chemical potentials $\mu_i$ are in general different. We assume that the reservoirs have a constant density of states. The dispersion relation is thus linear, $\epsilon_{\mathbf{k}}=v_F \left(\mathbf{k} - \mathbf{k}_F \right)$, with the Fermi velocity $v_F$ and Fermi momentum $\mathbf{k}_F$. We set $\hbar=1$ for simplicity.

The lattice is described by
\begin{equation}
H_{\text{chain}} = \epsilon \sum_{m = -l}^l d_m^{\dagger} d_m -\tau \sum_{m = -l}^{l - 1} \left( d_{m + 1}^{\dagger} d_m + \text{H.c.} \right),
\label{eq:chain_hamiltonian}
\end{equation}
where $d_m$ is the fermionic annihilation operator acting on site $m$ and $\tau$ is the tunneling amplitude within the lattice, with lattice spacing 1. The energy offset $\epsilon$ is equal for all sites. The chain length $M$ is chosen to be odd, $M=2l+1$, where $l$ is a non-negative integer, and $M=1$ corresponds to a single quantum dot. The second term in (\ref{eq:chain_hamiltonian}) only exists if the chain has more than one site. 
The Hamiltonian
$H_{\text{t}} = -\tau_1 \left[\psi^{\dagger}_{L}(\mathbf{0}) d_{-l} + d_l^{\dagger} \psi^{\phantom{\dagger}}_{R}(\mathbf{0}) + \text{h.c.}\right]$
describes tunneling between the ends of the chain and the respective reservoirs. 

We use the Keldysh technique~\cite{KamenevBook,SiebererDiehl2016} to describe the nonequilibrium situation with different chemical potentials of the reservoirs and a local particle loss. The Keldysh action
$S=\int d \omega/(2 \pi) \bar{\mathbf{\Psi}}(\omega) \mathcal{G}^{-1}(\omega) \mathbf{\Psi}(\omega)$
is written in the basis of fermionic coherent states parametrized by the Grassmann variables $\psi=(\psi^+, \psi^-)$. The vector elements correspond to the forward and backward time branches, which we rotate into $(\psi^1, \psi^2)$ using the bosonic convention. We use the basis
$\mathbf{\Psi} = 
\begin{pmatrix}
  \psi_L^1 &\psi_L^2	&d_{-l}^1	&d_{-l}^2
  &\dots&d_l^1	&d_l^2	&\psi_R^1 &\psi_R^2 
\end{pmatrix}^T$
to write the inverse Green's function in a tridiagonal block form
\begin{align}
\mathcal{G}^{-1} = 
\begin{pmatrix}
L	&T_1	&0		&		&\dots		&		&0	\\
T_1	&D_{-l}	&T		&0	\\
0	&T		&\ddots	&T	\\
	&0		&T		&D_0	&T			\\
\vdots	&	&		&T		&\ddots	&T	\\
	&		&		&		&T		&D_l	&T_1	\\	
0	&		&		&		&		&T_1	&R
\end{pmatrix}.
\label{eq:inverse_greens_function}
\end{align}

The corner blocks are $2\times 2$ matrices with the structure 
\begin{equation}
R/L=  
\begin{pmatrix}
0			&[\mathcal{G}_{L/R}^A]^{-1} \\
[\mathcal{G}_{L/R}^R]^{-1}	&[\mathcal{G}_{L/R}^K]^{-1}
\end{pmatrix},
\label{eq:leads}
\end{equation}
written in terms of the retarded (R), advanced (A), and Keldysh (K) Green's functions.
They correspond to the leads modeled by local Green's functions at $\mathbf{r}=0$ where the tunneling occurs:
\begin{align}
\mathcal{G}^{R, A}_{L/R}(\mathbf{r} = 0, \omega) = \frac{1}{\mathcal{V}} \sum_{\mathbf{k} = -\Lambda/v_F}^{\Lambda/v_F} \frac{1}{\omega - \epsilon_{\mathbf{k}} \pm i \eta}.
\label{eq:greens_function_r0}
\end{align}
Here, $\mathcal{V}$ is the volume of the reservoirs and $i\eta$ is an infinitesimal imaginary part. The Keldysh component is given by $\mathcal{G}_{L/R}^K=(\mathcal{G}_{L/R}^R - \mathcal{G}_{L/R}^A) \tanh \left[ (\omega - \mu_{L/R})/(2 T) \right]$ with temperature $T$~\cite{KamenevBook}. We consider here $T=0$. The reservoir eigenstates have a cutoff $\pm \Lambda/v_F$ as the linear dispersion relation is otherwise unbounded. While in the limit $\Lambda \to \infty$, the real part of (\ref{eq:greens_function_r0}) vanishes, keeping a finite real part is connected to the appearance of bound states outside the reservoir energy continuum~\cite{note_about_bound_states}.
The blocks $D_j$ correspond to the lattice sites and have the same structure as (\ref{eq:leads}),
\begin{equation}
D_{|j| > 1} =
  \begin{pmatrix}
0			&\omega - \epsilon - i \eta \\
\omega - \epsilon + i \eta	&2 i \eta \tanh \left( \frac{\omega - \epsilon}{2 T} \right)
  \end{pmatrix}.
\end{equation}
At the three central sites, the loss leads to a finite imaginary part (see SM~\cite{supplementary}) which modifies the matrix elements,
\begin{equation}
D_{j = 0, \pm 1} =
  \begin{pmatrix}
0			&\omega - \epsilon - i \gamma_j/2 \\
\omega - \epsilon + i \gamma_j/2	& i \gamma_j
  \end{pmatrix}.
\end{equation}
The dissipation rate on sites $j=\pm1$ is chosen symmetrically, $\gamma_1=\gamma_{-1}<\gamma_0$, to model for instance a dissipative laser beam with a Gaussian profile. We mostly set $\gamma_1=0$ but discuss briefly the consequences of nonzero $\gamma_1$. The off-diagonal blocks contain the tunneling matrix elements
\begin{equation*}
T_1 = 
\begin{pmatrix}
0	&\tau_1	\\
\tau_1	&0
\end{pmatrix},
\hspace{1cm}
T = 
\begin{pmatrix}
0	&\tau	\\
\tau	&0
\end{pmatrix}.
\end{equation*}

To characterize transport and the properties of the steady state, we calculate the current and the particle density distribution within the lattice. The conserved current is related to the change of particle number in the reservoirs (see~\cite{JinGiamarchi2020,supplementary} for details),
$I=-\frac{1}{2} (d/dt) \braket{N_L - N_R}$,
where $N_i=\int d\mathbf{r} \psi_i^{\dagger}(\mathbf{r}) \psi_i^{\dagger}(\mathbf{r})$ is the particle number operator. One can compute the full non-linear current-voltage characteristics from the above action but we focus 
on the conductance $G=\lim_{V \to 0} I/V$, where $V=\mu_L-\mu_R$ is the voltage, and fix $\mu_{L, R}=\pm V/2$. In natural units, the conductance quantum is $G_0=1/(2\pi)$.

\begin{figure}[h!]
\includegraphics[width=\linewidth]{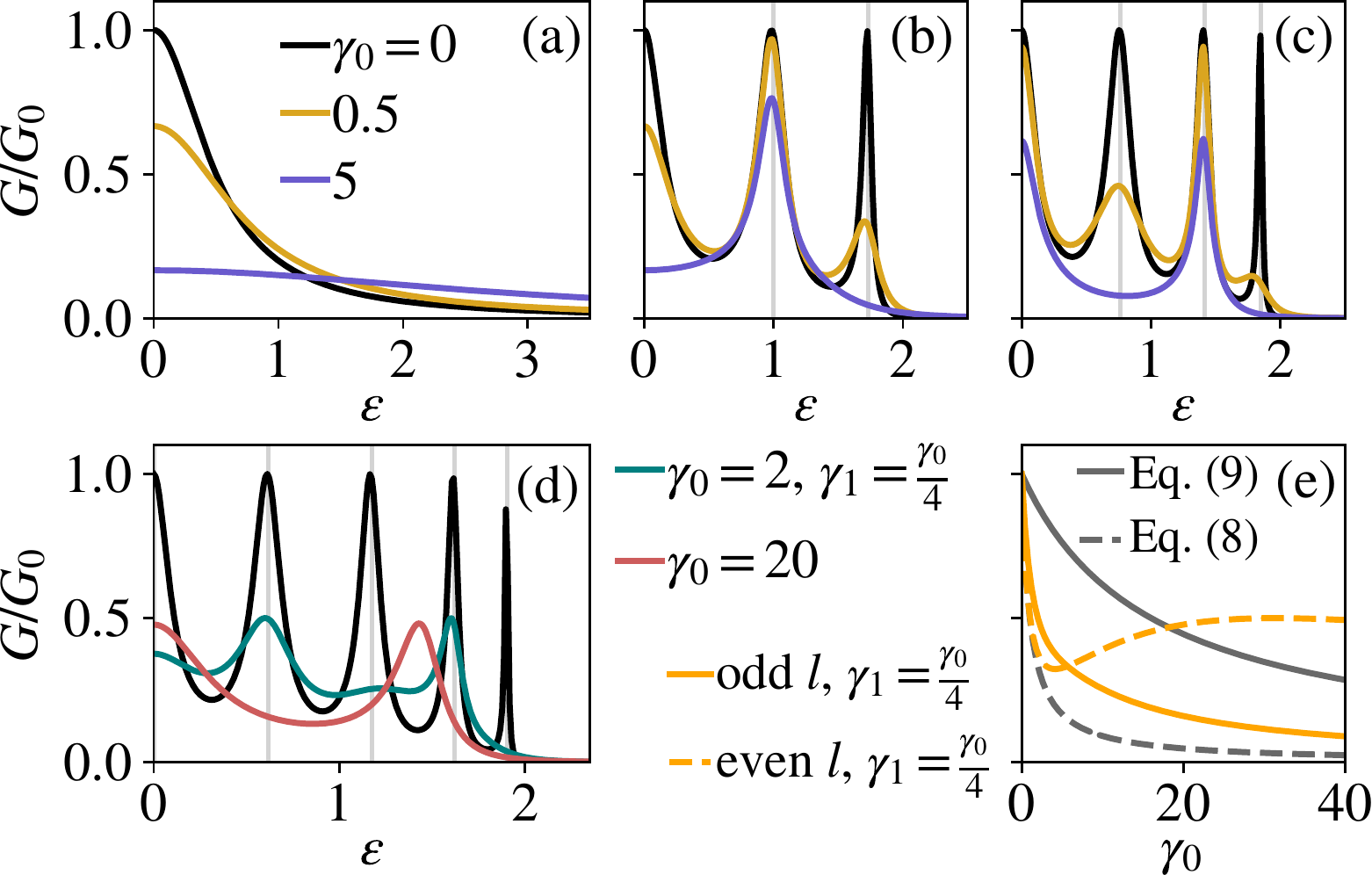}
\caption{(a) The conductance of a quantum dot as a function of the energy offset $\epsilon$, given by (\ref{eq:quantum_dot_conductance}), with $\tau=1$ and $\tau_1=0.5$ in units of $(\pi \rho_0)^{-1}$. In these units, $\tau_1$ has dimension $1/\sqrt{\mathcal{V}}$, while $\epsilon$, $\tau$, and $\gamma_j$ have dimension $1/\mathcal{V}$. The peak is reduced and broadened with $\gamma_0 > 0$. We plot only the $\epsilon>0$ region since $G(-\epsilon)=G(\epsilon)$. (b,c) The conductance of a lattice of $5$ and $7$ sites, respectively, with a loss only on the center site ($\gamma_1=0$). Vertical lines indicate the values of $\epsilon$ at which the single-particle eigenenergies are zero in the $\tau_1 \to 0$ limit. (d) For a nine-site lattice with $\gamma_1=\gamma_0/4$, resonance peaks at the center and close to the edges of the spectrum persist at large~$\gamma_0$. (e) The conductance as a function of $\gamma_0$ at the $\epsilon=0$ resonance. When $\gamma_1=0$, it is given by (\ref{eq:even_l}) and~(\ref{eq:odd_l}). For even $l$ ($1,5$, and $9$ sites), there is a faster decay than for odd $l$ (3 and 7 sites), explained by the spatial symmetry of the relevant eigenstates. For $\gamma_1 > 0$, a faster decay occurs for odd and a \emph{nonmonotonic} dependence for even~$l$ (see SM~\cite{supplementary}).}
\label{fig:resonances}
\end{figure}

In the case of a quantum dot coupled to reservoirs ($M=1$), only $\gamma_0$ is present. The conductance is
\begin{equation}
G_{\text{dot}} = \frac{1}{2 \pi} \frac{4 \Gamma (\gamma_0 + 4 \Gamma)}{4 \epsilon^2 + (\gamma_0 + 4 \Gamma)^2},
\label{eq:quantum_dot_conductance}
\end{equation}
where $\Gamma=\pi \rho_0 \tau_1^2$, and $\rho_0$ is the constant density of states per unit volume of the reservoirs. For $\gamma_0=0$, $G_{\textrm{dot}}$
is a Lorentzian function with width~$4\Gamma$.
Its maximum occurs when the energy level of the quantum dot $\epsilon$ coincides with the chemical potential in the reservoirs, here at $\epsilon=0$. At this value, the system is perfectly conducting with $G_{\text{dot}}=G_0$ and the conductance is independent of the tunneling $\tau_1$. A particle loss leads to a reduction of the maximum and a broadening of the Lorentzian peak, seen in Fig.~\ref{fig:resonances}(a).

For a lattice of $M$ sites, there are $M$ such resonances~\cite{JinGiamarchi2020} of the conductance, as seen for $M=5,7$, and~$9$ in Fig.~\ref{fig:resonances}(b--d). The horizontal axis is restricted to positive $\epsilon$ since $G(\epsilon)$ is an even function. To gain insight into the positions of the resonances, we consider the single-particle eigenstates of an isolated ($\tau_1=0$) lattice. 
There are $M$ eigenstates with eigenenergies symmetric around $\epsilon$. The resonances occur approximately when an eigenenergy $E_n$ coincides with the chemical potential in the reservoirs, here when $E_n=0$. 
The eigenenergies are indicated with vertical lines in Figs.~\ref{fig:resonances}(b--d). Whereas the number of maxima can be fully understood by this consideration, their positions are only exact in the $\tau_1 \to 0$ limit; the eigenenergies of the chain are shifted when it is coupled to reservoirs. This deviation is visible in Fig.~\ref{fig:resonances} where $\tau_1=0.5$. Quite remarkably, even if the maxima are shifted for $\tau_1 > 0$, the conductance at the maxima is perfect, $G/G_0=1$, in the absence of particle loss. This was checked for lattice sizes $M=3$ to $9$~\cite{supplementary}.

While a loss at the center site reduces the conductance peaks, every second peak is only very weakly reduced, as seen in Fig.~\ref{fig:resonances}(b,c). This interesting behavior stems from the fact that, for an isolated lattice,
half of the eigenstates are antisymmetric and have a node at the center where the particle loss takes place. Particles in antisymmetric eigenstates are therefore not depleted by the loss~\cite{WolffKollath2020}, and transport through these eigenstates -- at values of $\epsilon$ where an eigenenergy coincides with the chemical potential of the reservoirs -- is only weakly affected. Symmetric eigenstates on the other hand are depleted due to the nonzero overlap with the lossy site. This leads, as for the quantum dot, to a reduction of conductance. 
When an extended loss is present, as in Fig.~\ref{fig:resonances}(d), all eigenstates are depleted even in the $\tau_1 \to 0$ limit since they cannot have a node on three neighboring sites. However, for $\gamma_1 < \gamma_0$ and moderate values of $\gamma_0$, the maxima arising from symmetric eigenstates are still reduced more than the ones from antisymmetric eigenstates. Interestingly, for a lattice of nine sites, a larger $\gamma_0$ leads to a reinforcement of the peak at $\epsilon=0$. Resonances close to the edges of the spectrum are also preserved while the others are suppressed. The outermost resonances preserved at large $\gamma_0$ arise from eigenstates with a node at $j=0$ and only a small overlap with $j=\pm 1$.

We now analyze the conductance peak at $\epsilon=0$ in more detail for different lattice sizes.
The eigenstate in the middle of the spectrum is symmetric for even and antisymmetric for odd $l$. The conductance at $\epsilon=\gamma_1=0$ is
\begin{numcases}{G(\epsilon = 0) =}
\frac{1}{2 \pi} \frac{4 \Gamma}{\gamma_0 + 4 \Gamma}	&$l = 0, 2, 4$,
\label{eq:even_l} \\
\frac{1}{2 \pi} \frac{4 \tau^2}{\gamma_0 \Gamma + 4 \tau^2}	&$l = 1, 3$. 
\label{eq:odd_l}
\end{numcases}
The corresponding expressions are given in the SM~\cite{supplementary} for $\gamma_1 > 0$. Equations~(\ref{eq:even_l}) and~(\ref{eq:odd_l}) can be expanded at small $\gamma_0$ as
\begin{align}
\frac{4 \Gamma}{\gamma_0 + 4 \Gamma} \approx 1 - \frac{\tau}{4 \Gamma} \frac{\gamma_0}{\tau}, \hspace{5mm}
\frac{4 \tau^2}{\gamma_0 \Gamma + 4 \tau^2} \approx 1 - \frac{\Gamma}{4 \tau} \frac{\gamma_0}{\tau}.
\end{align}
When $\Gamma \ll \tau$, the slope $\tau/(4 \Gamma)$ is large compared to $\Gamma/(4 \tau)$, resulting in a larger reduction of the conductance with $\gamma_0$ for symmetric eigenstates [see Fig.~\ref{fig:resonances}(e)]. We relate the reduction for antisymmetric eigenstates to a symmetry breaking: the coupling to the reservoirs breaks the reflection symmetry around the lossy site, and the wavefunction gains a finite value there. When the loss extends to the neighboring sites, the wavefunction at those sites also plays a role. The antisymmetric eigenstates have a finite overlap with $j=\pm 1$, leading to a faster decay compared to $\gamma_1=0$. The symmetric eigenstates in contrast have a node at $j=\pm 1$. This leads to a nonmonotonic behavior: After an initial decay, the conductance is enhanced by the dissipation and even exceeds the value for a strictly local loss. In the $\gamma_0 \to \infty$ limit, it again decreases as~$\sim 1/\gamma_0$.

\begin{figure}[h!]
\includegraphics[width=\linewidth]{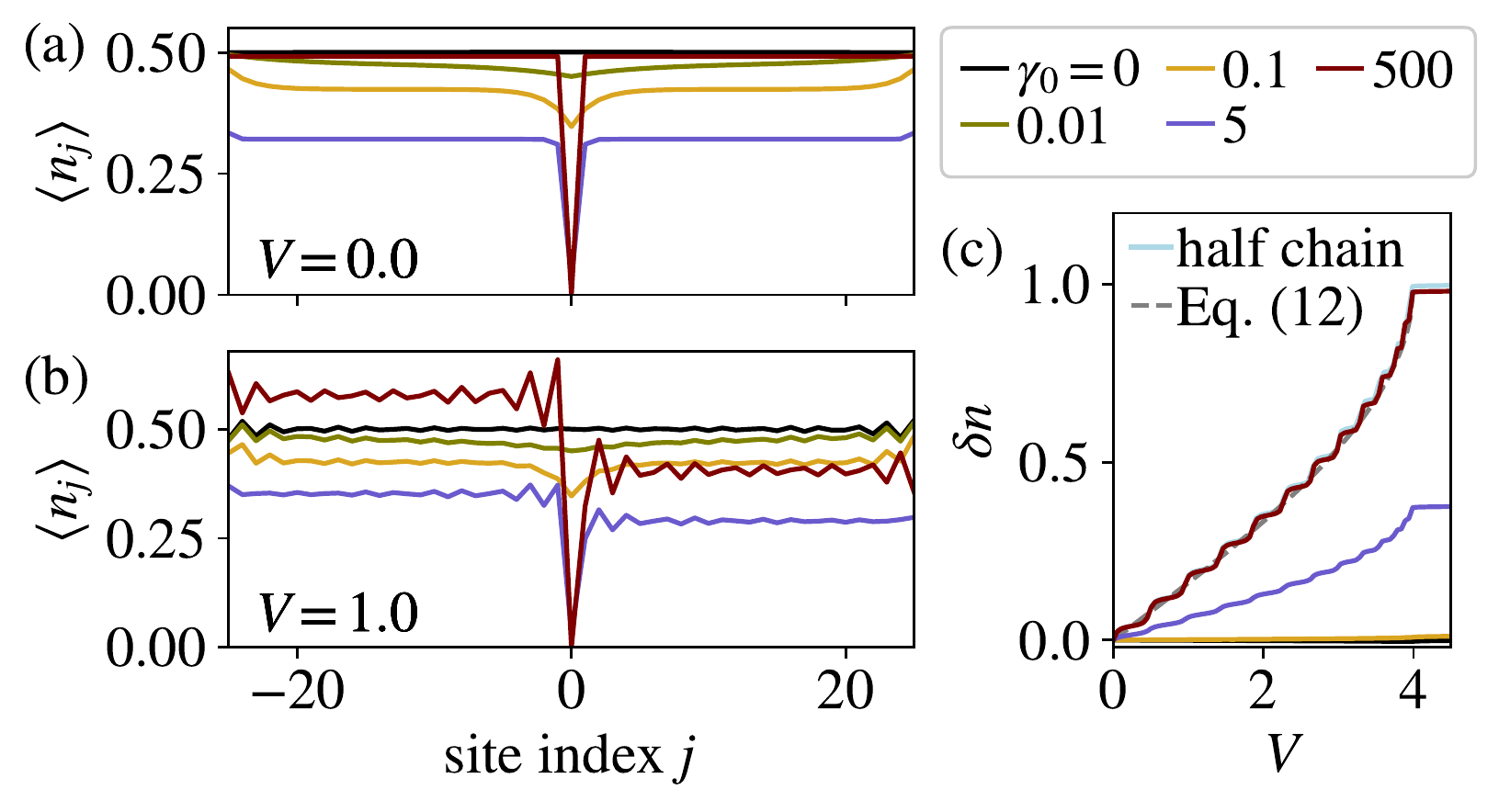}
\caption{(a, b) Particle density $\braket{n_j} = \braket{d_j^{\dagger} d_j^{\phantom{\dagger}}}$ in a 51-site lattice as a function of the position $j$, for different voltages $V$ and losses $\gamma_0$ with $\gamma_1=\epsilon=0$, $\tau=1$ and $\tau_1=0.5$. The parameters are in units of $(\pi \rho_0)^{-1}$ as in Fig.~\ref{fig:resonances}. A density imbalance between the left and right sides develops as a combined effect of the finite voltage and dissipation. (c) The average density imbalance $\delta n$, defined in (\ref{eq:average_imbalance}), as a function of voltage. We compare the largest dissipation rate $\gamma_0=500$ to a chain of $(M - 1)/2=25$ sites coupled to a single reservoir at equilibrium (``half chain''). The two results overlap and coincide with the imbalance determined by the reservoir density, given by~(\ref{eq:filling}).}
\label{fig:finite_voltage_L51}
\end{figure}

When the coupling~$\tau_1$ is finite, the particle loss leads to a reduction of conductance and non-ballistic transport. To further understand how the ballistic transport is altered, we analyze the steady-state particle density distribution, shown in Fig.~\ref{fig:finite_voltage_L51} for a lattice of 51 sites. We set $\gamma_1=0$ since the results are essentially the same for the extended loss. We focus on the interplay of the finite voltage and loss, but for completeness, panel~(a) shows the zero-voltage case. The setup in Refs.~\cite{FromlDiehl2019,FromlDiehl2020,MullerDiehl2021} is similar, except in our analysis, 
the coupling to the reservoirs is explicitly described. The loss leads to a density minimum at the lossy site while the density is nearly uniform in the surrounding lattice. This background density has a nonmonotonic dependence on $\gamma_{0}$ associated with the quantum Zeno effect~\cite{FromlDiehl2019,FromlDiehl2020}.

When a voltage is present but there is no loss, the density distribution is approximately uniform apart from Friedel oscillations (cf. Fig.~\ref{fig:finite_voltage_L51}(b), $\gamma_0=0$). Transport through the chain is ballistic and a voltage drop occurs only at the contacts. The situation changes drastically when particle losses are present. The average density becomes higher on the left side of the lossy site and lower on the right. 
The average density imbalance
\begin{equation}
\delta n = \frac{2}{L - 1} \left(\sum_{j<0} \braket{n_j} - \sum_{j>0} \braket{n_j} \right)
\label{eq:average_imbalance}
\end{equation}
is shown as a function of voltage in Fig.~\ref{fig:finite_voltage_L51}(c). 
It rises approximately linearly for small $V$ and saturates when the voltage exceeds the bandwidth $4 \tau$, as there are no eigenstates which could be filled by chemical potentials outside the lattice energy band. The slope and saturation value of $\delta n(V)$ depend strongly on the dissipation strength. For small $\gamma_0$, hardly any imbalance develops, and transport through the chain is almost ballistic. In contrast, for large $\gamma_0$, 
a large imbalance arises, reaching the maximum value 1 for $\gamma_0 \to \infty$ and $V > 4 \tau$. In the presence of the loss, transport is thus no longer perfectly ballistic. This agrees with the reduction of conductance for increasing loss. 

For $\gamma_0 \to \infty$, the lossy site is almost completely emptied, effectively cutting the chain in two. The average density on either side approaches that of a chain of half the length, coupled to a single reservoir with the chemical potential $\mu_{L/R}$ at equilibrium.
Figure~\ref{fig:finite_voltage_L51}(c) shows that the average imbalance for $\gamma_0=500$ indeed overlaps with the result for the half chain. Furthermore, the slope of $\delta n(V)$ can be estimated from
the particle density in the reservoir coupled to each half, $\braket{n_{\text{res}}^{L, R}}$, by equating the chemical potential $\pm V/2$ with the energy $\epsilon_{k_F}=-2 \tau \cos(k_F)$ -- the lattice dispersion relation -- where $k_F=\pi \braket{n_{\text{res}}}$. The filling factor is then given by 
\begin{equation}
\braket{n_{\text{res}}^{L, R}} = \frac{1}{\pi}\arccos \left(\mp \frac{V}{4 \tau} \right).
\label{eq:filling}
\end{equation}
The resulting difference $\braket{n_{\text{res}}^L} - \braket{n_{\text{res}}^R}$ agrees very well with the result for $\gamma_0=500$ and the one for a half chain. 
We further observe that at large $\gamma_0$, the average densities develop a step-like substructure. These discrete changes in density result from adding (on the left) or removing (on the right) particles one by one when the chemical potential of the reservoirs changes. Indeed, for $25$ sites on either side, a change from half to full filling on the left and to an empty lattice on the right corresponds to 12.5 particles, which equals the number of steps in Fig.~\ref{fig:finite_voltage_L51}(c). For larger lattices, these steps become more frequent, suggesting that the density approaches the smooth function~(\ref{eq:filling}) in the $M \to \infty$ limit. 

In summary, we show that particle transport can be very differently affected by a local loss depending on the underlying symmetry of the eigenstate responsible for the transport. In particular, conductance can remain nearly perfect despite the loss. In the nonlinear-response regime with a finite voltage, we find novel dissipative steady states characterized by a density drop at the lossy site. These phenomena could be observed in cold-atom experiments, where local particle losses have already been explored. Extending the analysis to bosonic~\cite{LabouvieOtt2016} or interacting systems is an exciting outlook. An interesting question is what kind of transport properties and nonequilibrium steady states arise in the presence of losses when the system is for example in a correlated insulator~\cite{LebratEsslinger2018} or superfluid~\cite{HusmannBrantut2015} state, or when a local loss leads to long-range coherence~\cite{DuttaCooper2020}.
Our analysis focuses on steady-state behavior, and an important question is how, and on what timescale, the steady state is reached. This long-time evolution is not easily accessed by our current approach and is a considerable challenge for future research. However, dissipative steady states have been realized in experiments with local losses~\cite{BarontiniOtt2013, CormanEsslinger2019}, and we therefore expect that the properties of the steady states studied here could be observed on experimentally relevant timescales.

\begin{acknowledgments}

We thank C. Berthod, S. Diehl, T. Esslinger, P. Fabritius, M.-Z. Huang, J. Mohan, H. Ott, M. Talebi, S. Uchino, and S. Wili for helpful and inspiring discussions.
We acknowledge funding from the Deutsche Forschungsgemeinschaft (DFG, German Research Foundation) in particular under project number 277625399 - TRR 185 (B3) and project number 277146847 - CRC 1238 (C05), Einzelantrag KO 4771/2-1 and under Germany’s Excellence Strategy – Cluster of Excellence Matter and Light for Quantum Computing (ML4Q) EXC 2004/1 – 390534769 and the European Research Council (ERC) under the Horizon 2020 research and innovation programme, grant agreement No.~648166 (Phonton). This work was supported in part by the Swiss National Science Foundation under Division II. 
\end{acknowledgments}

\bibliographystyle{apsrev4-1-with-titles}
\bibliography{references_Bonn.bib, bibfile}

\begin{thebibliography}{40}
\bibitem [{not()}]{note_about_bound_states}%
  \BibitemOpen
  \href@noop {} {}\bibinfo {note} {{When the cutoff $\Lambda$ is finite, bound
  states outside the reservoir energy continuum contribute to the particle
  densities at the edge sites. The bound-state contribution decays with
  $\Lambda$, and we use here values for which it is negligible.}}\BibitemShut
  {Stop}%
\end{thebibliography}

\end{document}


\title{Symmetry-protected transport through a lattice with a local particle loss - supplementary material}

\maketitle

\section{Observables}
\label{sec:observables}

We consider an open quantum system described by the quantum master equation
\begin{equation}
\frac{d\rho}{dt} = -i [H, \rho] + \sum_{j = -1}^1 \gamma_j \left[ L_j \rho L_j^{\dagger} - \frac{1}{2} \left\{ L_j^{\dagger} L_j, \rho \right\} \right],
\label{eq:master_equation}
\end{equation}
where $\rho$ is the density operator and $H$ the Hamiltonian. The Lindblad operator $L_j$ is here the annihilation operator $d_j$ at sites $j=0,\pm 1$. 
To characterize transport, we calculate the particle current $I$ through the lattice system. The current is connected to the change of particle numbers in the reservoirs, 
\begin{equation}
I = -\frac{1}{2} \frac{d}{dt}\braket{N_L - N_R},
\label{eq:conserved_current}
\end{equation}
where the expectation value is defined as $\braket{A} = \text{Tr}(A \rho)$ for a generic operator $A$. The particle number operator is $N_i = \int d\mathbf{r} \psi_i^{\dagger}(\mathbf{r}) \psi_i^{\phantom{\dagger}}(\mathbf{r})$ with $i = L, R$. We obtain its time derivative as
\begin{align}
\frac{d}{dt} \braket{N_i} &= \frac{d}{dt} \text{Tr}\left( N_i \rho(t) \right) \\
&= -i\text{Tr}\left( N_i [H, \rho] \right) \\
&\hspace{1mm}+ \sum_{j = -1}^1 \gamma_j \text{Tr}\left( N_i d_j \rho d_j^{\dagger} - \frac{1}{2} N_i \left\{ d_j^{\dagger} d_j, \rho \right\} \right) 
\label{eq:particle_loss_term} \\
&= -i \braket{[N_i, H_t]},
\notag 
\end{align}
where term~(\ref{eq:particle_loss_term}) is zero as $[N_i, d_j] = 0$. On the last line, we observe that $N_i$ commutes with all terms in the Hamiltonian apart from $H_t$.
Thus, the time derivatives entering Eq.~(\ref{eq:conserved_current}) are found as 
\begin{align}
\begin{split}
\frac{d}{dt}\braket{N_L} &= i \tau_1 \left( \braket{\psi_L^{\dagger}(\mathbf{0}) d_{-l}} - \braket{d_{-l}^{\dagger} \psi_L(\mathbf{0})} \right), \\
\frac{d}{dt} \braket{N_R} &= i \tau_1 \left( \braket{\psi_R^{\dagger}(\mathbf{0}) d_{l}} - \braket{d_{l}^{\dagger} \psi_R(\mathbf{0})} \right).
\end{split}
\label{eq:time_derivatives}
\end{align}
The lattice size is denoted by $M = 2 l + 1$. The dissipation strength $\gamma_{0, \pm 1}$ does not appear explicitly in the expression for the current since the particle loss operator $d_j$ commutes with $N_i$.

\section{Keldysh formalism}

\subsection{Keldysh action for the dissipative system}

We compute the nonequilibrium expectation values of Eq.~(\ref{eq:time_derivatives}) as path integrals over a closed time contour. As we are interested in steady-state expectation values, the integration contour extends from $t \to -\infty$ to $\infty$ and back. One formulates the nonequilibrium action in terms of Grassmann fields $\psi^{\pm}, \bar{\psi}^{\pm}$ for the forward ($+$) and backward ($-$) time contours. The action of the open quantum system is~\cite{SiebererDiehl2016}
\begin{align}
\begin{split}
S = \int_{-\infty}^{\infty} dt \Big[ &\bar{\psi}^+ i \partial_t \psi^+ - \bar{\psi}^- i \partial_t \psi^- \\
&- i \mathcal{L}(\bar{\psi}^+, \psi^+, \bar{\psi}^-, \psi^-) \Big].
\end{split}
\end{align}
In the case of the lattice coupled to leads, $\psi^{\pm}$ denotes the vector $\psi^{\pm} = (\psi_L^{\pm} \; d_{-l}^{\pm} \; \dots \; d_0^{\pm} \; \dots \; d_l^{\pm} \; \psi_R^{\pm})^T$. In writing the fermionic Keldysh action, one has to take into account the anticommutation relations of the Grassmann variables~\cite{KamenevBook}. The Lindblad term $\mathcal{L}$ is
\begin{align}
\begin{split}
\mathcal{L} (&\bar{\psi}^+, \psi^+, \bar{\psi}^-, \psi^-) = -i (H^+ - H^-) \\
&+\sum_{j = -1}^1 \gamma_j \Big[ \bar{d}_j^- d_j^+ - \frac{1}{2} \bar{d}_j^+ d_j^+ 
- \frac{1}{2} \bar{d}_j^- d_j^- \Big],
\end{split}
\label{eq:lindblad_term}
\end{align}
where $H^{\pm}$ is a function of $\psi^{\pm}$. The first term on the right-hand side represents the unitary time evolution by the Hamiltonian. The terms on the second line, proportional to $\gamma_j$, describe the dissipative process. 

The action in total can now be written as a sum of the coherent and dissipative terms, 
\begin{equation}
S = \sum_{i = L, R} S_i + S_{\text{tun}} + S_{\text{chain}} + S_{\text{loss}}. 
\label{eq:action_sum}
\end{equation}
The coherent terms are the first three,
\begin{align*}
&S_i = \int_{-\infty}^{\infty} dt \left( \bar{\psi}_i^+ i \partial_t \psi_i^+ - \bar{\psi}_i^- i \partial_t \psi_i^- - H_i^+ + H_i^- \right), \\
&S_{\text{tun}} = \tau_1 \int_{-\infty}^{\infty} dt \Big( \bar{d}_{-l}^+ \psi_L^+ + \bar{\psi}_L^+ d_{-l}^+ + \bar{d}_{l}^+ \psi_R^+ + \bar{\psi}_R^+ d_{l}^+ \\
&\hspace{2cm} - \bar{d}_{-l}^- \psi_L^- - \bar{\psi}_L^- d_{-l}^- - \bar{d}_{l}^- \psi_R^- - \bar{\psi}_R^- d_{l}^- \Big), \\
&S_{\text{chain}} = \sum_{j = -l}^{l} S_j + \tau \sum_{j = -l}^{l - 1} \int_{-\infty}^{\infty} dt \Big[ \bar{d}_{j + 1}^+ d_j^+ + \bar{d}_j^+ d_{j + 1}^+ \\
&\hspace{4cm} - \bar{d}_{j + 1}^- d_j^- - \bar{d}_j^- d_{j + 1}^- \Big],
\end{align*}
where the term $S_j$ for each lattice site is
\begin{equation}
S_j = \int_{-\infty}^{\infty} dt \Big[ \bar{d}_j^+ \left( i \partial_t - \epsilon \right) d_j^+ 
- \bar{d}_j^- \left(i \partial_t - \epsilon \right) d_j^- \Big].
\end{equation}
We perform the Keldysh rotation as
\begin{align}
\psi^1 = \frac{1}{\sqrt{2}} \left( \psi^+ + \psi^- \right), \hspace{5mm}
\psi^2 = \frac{1}{\sqrt{2}} \left( \psi^+ - \psi^- \right),
\label{eq:keldysh_rotation}
\end{align}
using the boson notation for fermionic coherent states~\cite{KamenevBook} rather that the usual one of Larkin and Ovchinnikov. In the new basis, $S_i$ and $S_j$ are represented in matrix form as 
\begin{equation}
\mathcal{S}_x = \int \frac{d \omega}{2 \pi} 
\begin{pmatrix}
\bar{\psi}^1 &\bar{\psi}^2
\end{pmatrix}
\mathcal{G}_x^{-1}(\omega)
\begin{pmatrix}
\psi^1	\\
\psi^2
\end{pmatrix},
\label{eq:action_matrix}
\end{equation} 
where $x = i, j$, transformed into frequency basis. The inverse Green's function $\mathcal{G}_x^{-1}$ has the standard structure
\begin{equation}
\mathcal{G}_x^{-1}(\omega) = 
\begin{pmatrix}
0	&\left[ \mathcal{G}_x^{\mathcal{A}} \right]^{-1} \\
\left[ \mathcal{G}_x^{\mathcal{R}} \right]^{-1}	&\left[ \mathcal{G}_x^{-1}\right]^{\mathcal{K}}
\end{pmatrix},
\label{eq:general_inverse_greens_function}
\end{equation}
where $\left[ \mathcal{G}_x^{-1}\right]^{\mathcal{K}} = -\left[ \mathcal{G}_x^{\mathcal{R}} \right]^{-1} \mathcal{G}_x^{\mathcal{K}} \left[ \mathcal{G}_x^{\mathcal{A}} \right]^{-1}$, and $\mathcal{G}_x^{\mathcal{A}}$, $\mathcal{G}_x^{\mathcal{R}}$, and $\mathcal{G}_x^{\mathcal{K}}$ are the advanced, retarded, and Keldysh components. 
The reservoirs and lattice sites apart from $j = 0$ evolve unitarily in time. In this case, one obtains $\mathcal{G}_x^{\mathcal{K}}$ as
\begin{equation}
\mathcal{G}_x^{\mathcal{K}} = (\mathcal{G}_x^{\mathcal{R}} - \mathcal{G}_x^{\mathcal{A}}) [1 - 2 n_F(\omega - \mu_x)],
\label{eq:fluctuation-dissipation}
\end{equation} 
where $n_F(\omega) = (e^{\omega/T} + 1)^{-1}$ with temperature $T$ denotes the Fermi-Dirac distribution. For the lattice sites, $\mu_j = \epsilon$.
The $S_j$ terms for the three central sites are modified by the addition of $S_{\text{loss}}$, given by the second term of Eq.~(\ref{eq:lindblad_term}). Applying Eq.~(\ref{eq:keldysh_rotation}), we get
\begin{align*}
S_{\text{loss}} = \sum_{j = -1}^1 \int_{-\infty}^{\infty} \frac{d\omega}{2 \pi} 
\begin{pmatrix}
\bar{d}_j^1 &\bar{d}_j^2
\end{pmatrix}
\begin{pmatrix}
0	&-\frac{i \gamma_j}{2} \\
\frac{i \gamma_j}{2}	&i\gamma_j
\end{pmatrix}
\begin{pmatrix}
d_j^1	\\
d_j^2
\end{pmatrix}.
\end{align*}

\subsection{Correlation functions as matrix elements}

Two-operator expectation values are calculated as Gaussian path integrals
\begin{equation}
\braket{\psi^a \bar{\psi}^b} = \int \mathcal{D}[\bar{\psi}, \psi] \psi^a \bar{\psi}^{b} e^{i S[\bar{\psi}, \psi]} = i \mathcal{G}_{a b},
\label{eq:correlation_function}
\end{equation}
where $a, b \in \{ 1, 2 \}$. Here, $S$ denotes the total action~(\ref{eq:action_sum}), and the corresponding inverse Green's function $\mathcal{G}^{-1}$ is given by Eq.~(2) of the main text. The matrix elements $\mathcal{G}_{a b}$ are found by inverting $\mathcal{G}^{-1}$.
The expectation value of the current $I$ of Eq.~(\ref{eq:conserved_current}) is written in the Keldysh representation as
\begin{align*}
I &= \frac{i \tau_1}{4} \int_{-\infty}^{\infty}\frac{d \omega}{2 \pi} \Big( \braket{d_{-l}^1 \bar{\psi}_L^1(\mathbf{0})} - \braket{\psi_L^1(\mathbf{0}) \bar{d}^1_{-l}}  \\
&\hspace{3cm}+ \braket{\psi_R^1(\mathbf{0}) \bar{d}_l^1} - \braket{d_l^1 \bar{\psi}_R^1(\mathbf{0})} \Big) \\
&= \frac{\tau_1}{4} \int_{-\infty}^{\infty}\frac{d \omega}{2 \pi} (\mathcal{G}_{3, 1} - \mathcal{G}_{1, 3} + \mathcal{G}_{2 M + 3, 2 M + 1} \\
&\hspace{5cm}- \mathcal{G}_{2 M + 1, 2 M + 3} ),
\end{align*} 
and the particle density in the lattice is given by
\begin{equation*}
\begin{split}
\braket{n_j} &= \int_{-\infty}^{\infty}\frac{d \omega}{2 \pi} \frac{1}{2} \left( \braket{d_j^1 \bar{d}_j^1} - \braket{d_j^1 \bar{d}_j^2} + \braket{d_j^2 \bar{d}_j^1} \right) \\
&= \int_{-\infty}^{\infty}\frac{d \omega}{2 \pi} \frac{1}{2} \Big( \mathcal{G}_{2(l + j) + 3, 2(l + j) + 3} \\
&\hspace{5mm} - \mathcal{G}_{2(l + j) + 3, 2(l + j) + 4} + \mathcal{G}_{2(l + j) + 4, 2(l + j) + 3} \Big).
\end{split}
\end{equation*}

\section{Value of the conductance maxima}

We observe that the conductance as a function of the energy level $\epsilon$ of the lattice sites has maxima at positions which correspond approximately to resonances of the eigenstate energies. In the limit $\tau_1 \to 0$, the positions of the resonances are the values of $\epsilon$ for which the eigenstate energies of an isolated lattice coincide with the chemical potential in the reservoirs. The positions are shifted when $\tau_1 > 0$. We however observe that in the absence of dissipation, the value of the conductance at the maxima, such as in Fig.~2 of the main text, remains at $G/G_0 = 1$. For a lattice of three sites, at $\gamma_0 = \gamma_1 = 0$, the conductance has the expression
\begin{equation}
G = \frac{1}{2 \pi} \frac{4 \Gamma^2 \tau^4}{(\epsilon^2 + \Gamma^2) \left[ (\epsilon^2 - 2 \tau^2)^2 + \epsilon^2 \Gamma^2\right]}.
\label{eq:three_sites_conductance}
\end{equation}
By differentiating with respect to $\epsilon$, we get the positions of the maxima as
$\epsilon = 0$, $\epsilon = \pm \sqrt{2 \tau^2 - \Gamma^2}$, and G of Eq.~(\ref{eq:three_sites_conductance}) is 1 at these values. The same analysis for 5, 7, and 9 sites shows that $G/G_0 = 1$ at all maxima.

To model a local loss which extends to the sites next to the central one, we set $\gamma_1 = \gamma_{-1} > 0$. We find the conductance at $\epsilon = 0$ as
\begin{align}
G = \frac{1}{2 \pi} \frac{4 \Gamma \tau^2 \left( 4 \tau^2 + \gamma_0 \gamma_1 \right)}{\left(2 \tau^2 + \Gamma \gamma_1 \right) \left( 8 \Gamma \tau^2 + 2 \gamma_0 \tau^2 + \Gamma \gamma_0 \gamma_1 \right)}
\label{eq:extended_loss_even}
\end{align}
for $l = 2, 4$; $M = 5, 9$ and 
\begin{align}
G = \frac{1}{2 \pi} \frac{4 \Gamma \left(4 \tau^2 + \gamma_0 \gamma_1 \right)}{\left( 2 \Gamma + \gamma_1 \right) \left( 8 \tau^2 + 2 \Gamma \gamma_0 + \gamma_0 \gamma_1 \right)}
\label{eq:extended_loss_odd}
\end{align}
for $l = 1, 3$; $M = 3, 7$.

\bibliographystyle{apsrev4-1-with-titles}
\bibliography{references_Bonn.bib}